\documentclass{aastex}
\begin{document}
\title{Mars Odyssey Joins The Third Interplanetary Network}
\author{K. Hurley}
\affil{University of California, Berkeley, Space Sciences Laboratory, Berkeley, CA 94720-7450}
\email{khurley@ssl.berkeley.edu}
\author{I. Mitrofanov, A. Kozyrev, M. Litvak, A. Sanin, V. Grinkov and S. Charyshnikov}
\affil{Institute for Space Research, Moscow, Russia}
\author{W. Boynton, C. Fellows, K. Harshman, D. Hamara, and C. Shinohara}
\affil{Lunar and Planetary Laboratory, University of Arizona, Tempe AZ}
\author{R. Starr}
\affil{The Catholic University of America, Department of Physics, Washington DC 20064}
\author{T. Cline}
\affil{NASA Goddard Space Flight Center, Code 661, Greenbelt, MD 20771}
\begin{abstract}

The \it Mars Odyssey \rm spacecraft carries two experiments which are capable of detecting
cosmic gamma-ray bursts and soft gamma repeaters.  Since April 2001 they have
detected over 275 bursts and, in conjunction with the other spacecraft
of the interplanetary network, localized many of them rapidly and precisely enough to allow
sensitive multi-wavelength counterpart searches.  We present the \it Mars Odyssey \rm
mission and describe the burst capabilities of the two experiments in detail.  We explain how the spacecraft
timing and ephemeris have been verified in-flight using bursts from objects whose
precise positions are known by other means.  Finally, we 
show several examples of localizations and discuss future
plans for the \it Odyssey \rm mission and the network as a whole. 

\end{abstract}


\section{Introduction}

Interplanetary networks (IPNs) have played an important role in the studies of both
cosmic gamma-ray bursts (GRBs) and soft gamma repeaters (SGRs) for over two
decades.  Indeed, until the launch of \it BeppoSAX \rm in 1996, the only way to
derive arcminute positions for these objects was by comparing their arrival times at
distant spacecraft.  The current (third) IPN was formed when the \it Ulysses \rm
spacecraft was launched in 1990.  Over 25 spacecraft have participated in
the IPN since then, and the latest interplanetary mission to join the network
is \it Mars Odyssey \rm.  It seems fitting that this spacecraft should belong to the IPN,
since ``Odyssey'' and ``Ulysses'' both refer to the same saga of distant voyages.
Today, the IPN comprises the \it Ulysses, Konus-Wind, Ramaty High Energy Solar
Spectroscopic Imager \rm (RHESSI)\it, High Energy Transient Explorer
\rm(HETE)\it, Swift, \rm and \it Mars Odyssey \rm (MO)
missions and experiments, and, with a detection rate of ~200 events/year, 
is responsible for most GRB and SGR detections and localizations.  As a distant point in
the network, MO plays a crucial role: without it, only localizations to annuli or large error boxes would
be possible.  The triangulation, or arrival-time analysis method for localizing bursts has been
presented elsewhere (Hurley et al. 1999a,b).  In this paper, we concentrate on the properties of the two
MO experiments which make burst detection possible.  We note that this is the fifth attempt,
and the first successful one, to place a GRB detector in Mars orbit; the four previous attempts,
aboard the \it Phobos 1 \& 2 \rm (Sagdeev and Zakharov 1990) \it , Mars Observer \rm
(Metzger et al. 1992), and  \it Mars '96 \rm (Ziock et al. 1997) missions, met with limited or
no success due to mission failures.

\section{The \it Mars Odyssey \bf Mission}

The \it Mars Odyssey \rm mission is an orbiter whose objective is to provide a better understanding
of the climate and geologic history of Mars.  It was launched on 2001 April 7, and after a 6 month cruise phase,
reached Mars on 2001 October 24.  The mission then entered an aerobraking phase to circularize the orbit 
which lasted until 2002 January.  At the end of this phase, the spacecraft was orbiting
the planet every 1.964 h at an altitude between 370 and 432 km (Saunders et al. 2004).  The prime scientific mission 
then commenced, and at present, \it Odyssey \rm is in its first extended mission, which will continue
through 2006 September; a mission extension beyond that date is likely.

The spacecraft is shown in figure 1.  In its present orbit, Mars subtends approximately 27\% of the sky
(62 $\arcdeg$ half-angle) at the Odyssey spacecraft.  In general, the instruments are powered on continuously, and almost 100\% of 
the data is downlinked through the Deep Space Network during daily tracking passes.  A more complete
description of the mission has appeared in Saunders et al. (2004).
 
\subsection{The Gamma Sensor Head and the High Energy Neutron Detector}

The Gamma-Ray Spectrometer (GRS) is an instrument suite which includes two detectors
with GRB detection capabilities, the gamma sensor head (GSH), and the High Energy Neutron
Detector (HEND).
The principal objective of the GRS experiment is the determination of the
elemental abundances on Mars.  The GSH consists of a 6.7 cm diameter $\times$ 6.7 cm high 
(cross sectional area to GRBs $\sim$ 40 cm$^{2}$) right circular
cylindrical germanium detector which is passively cooled and mounted on a boom extending
6 m from the body of the spacecraft.  It records energy spectra between
$\sim$ 50 keV and 10 MeV
in a low time resolution mode ($\sim$20 s) until triggered by a burst.  
It then records GRB time histories
in a single energy channel with 32 ms resolution for 19.75 s, and can
retrigger immediately thereafter. The boom extension and detector cooling did not take place until after the end of the
aerobraking phase, and thus the experiment did not collect useful GRB data until then.  The in-orbit
background rate is ~100 c/s in the GRB energy channel, but it undergoes variations
due to numerous causes.  In order of decreasing importance, these
are a) the albedo from the cosmic gamma-ray background from the Martian surface, which
is different for different regions of the planet, b) seasonal changes on $\sim$ month
timescales such as CO$_2$ condensation in the polar cap regions (which
suppresses the upcoming gamma-radiation from the surface of the planet), and c) solar
proton events.  The GSH is
shown in figure 2.  More details may be found in Boynton et al. (2004).

The burst detection portion of the HEND experiment is 
based on two optically separate scintillation detectors (figure 3).  The first is a cylindrical stilbene
crystal with a diameter of 30 mm and a height of 10 mm, which is used for the detection of
high energy neutrons, and records gamma-rays as a by-product.  These counts
are measured continuously in the 350-3000 keV range with a time resolution of 1 second.
The second detector is a cylindrical CsI(Tl) anti-coincidence well surrounding the stilbene,
whose thickness is 10 mm, whose outer diameter is 50 mm, and whose height is 49 mm.  Thus its 
cross-sectional area to GRBs varies between $19.6$ cm$^2$ (on-axis), $24.5$ cm$^{2}$
(90 $\arcdeg$ off-axis), and $7.1$ cm$^{2}$ (180 $\arcdeg$ off-axis).  
In triggered mode, counts from the CsI are recorded in the $\sim$
30-1300 keV energy range with a time resolution of 250 ms,
and these data are used for triangulation.  The upper and lower energy limits
are only approximate, since the light collection in the cylindrical crystal depends 
upon the photon interaction point.  Energy spectra are not transmitted.  
The capacity of the counters is limited to 511 counts every 250 ms, so very strong
bursts can temporarily saturate the detector.  HEND is mounted on the body 
of the spacecraft.  The in-orbit background rates are $\sim 5$ counts/sec for the inner stilbene detector,
and $\sim 130$ counts/sec for the CsI anticoincidence.  
Both these rates undergo variations for the same reasons as the GSH, and in addition, 
because of HEND's lower energy threshold, due to trapped particles.  For example, for a period
of approximately 7 months starting on November 11, 2002, the background variations along the
orbit increased from their nominal value of about a factor of 2 to a factor of 30.
The cause of this increase is suspected to be trapped particles.  During this time, the duty cycle for
GRB detection decreased by a factor of 3.4.  These large variations disappeared around
June 9, 2003. Occasionally they reappear rather reduced in strength.  A background 
plot is shown in figure 4.  
More details of the instrument may be found in Boynton et al. (2004).

\section{Burst Detection and Statistics}

Bursts are detected in HEND and the GSH in different ways.  HEND is equipped with an
on-board trigger, but the telemetry allocation made it possible to transmit triggered data continuously.
Thus the trigger level was set to a value below the background level, and in effect, the
experiment triggers continuously and produces continuous data with 250 ms time resolution.
Searches for bursts in the HEND data may therefore be initiated by detections on other
IPN spacecraft, and may be carried out in the ground analysis.  Every candidate event detected by an IPN spacecraft
initiates such a search, and almost all the bursts detected by HEND to date have been found by searching
the data in the crossing time window which corresponds to these candidate events.
(Thus, for example, in the case of an event detected by \it Ulysses \rm,
the crossing time window is the Mars-Ulysses distance expressed in light-seconds;
the maximum distance is $\sim$ 3000 light-seconds.)  Up to now,
no exhaustive automatic ``blind'' search for bursts has been carried out in the HEND data, although
an algorithm to do this is being developed and tested.  GSH burst data, on
the other hand, come only from on-board triggers; each trigger initiates a search for a 
confirming event in the
data of the other IPN spacecraft in the appropriate crossing time windows.  
As high time resolution GSH data are only available for
triggered events, it is usually not practical to search the data for bursts detected by other
spacecraft.   

HEND was turned on during the cruise phase 
and detected its first confirmed GRB on 2001 May 8.  In the 1520 days
which followed, HEND and/or
GRS have detected $>$ 275 GRBs or SGRs which have been confirmed by detection aboard at least one other IPN spacecraft.
Thus the average burst detection rate during the cruise and orbital phases is about one burst every 
5.9 d.  This number is an average over the entire period, and does not take high background
periods, or times when the experiments were turned off, into account.  The true rate
would therefore be higher.  Initially, the spacecraft timing was not determined accurately
during the cruise phase and no triangulation results were announced.  
(These data have since been reprocessed for accurate timing.)  
Thus the \it Odyssey \rm mission became most useful to the IPN when it entered its orbital phase; 
at that point its separation from Earth was large enough for the triangulation method to produce small error boxes,
and the timing was known to good accuracy in real time.
The typical IPN annulus width as a function of time over the mission is shown in figure 5.

The GRB detection rate by 3 widely separated spacecraft (that is, \it Mars Odyssey,
Ulysses\rm, and any near-Earth mission) is about 1 every 7 days.  It is interesting to 
note that this rate is identical to that which was obtained when
the \it Near Earth Asteroid Rendezvous \rm (NEAR) mission was the other distant spacecraft
in addition to \it Ulysses \rm.  These numbers do not depend strongly
upon any comparison of the MO and NEAR instrument sensitivities, for the following
reason.  Due to its
small size, \it Ulysses \rm tends to be the least sensitive detector in the network.
We are counting only those bursts which were detected by \it Ulysses\rm, a
near-Earth spacecraft, and another, distant spacecraft.  As long as the sensitivities of
the near-Earth and the distant spacecraft are greater than that of \it Ulysses \rm, the
burst rate is determined primarily by \it Ulysses' \rm sensitivity, rather than by that
of the other detectors.

Figure 6 presents the distribution of the angles between
the HEND detector axis, which is always pointed towards Mars in the orbital phase of the
mission, and the GRBs detected by it.  This distribution agrees with the wide
field of view of the CsI anticoincidence, which is limited mainly by the Mars horizon;
the body of the spacecraft blocks a small fraction of the sky with relatively low atomic number
materials.

Figure 7 compares the count rates for events detected by HEND and \it
Ulysses \rm.  Since the HEND count rate is roughly proportional to that of \it Ulysses \rm,
this relation can be used for an approximate estimate of the fluence of HEND bursts.
The weakest confirmed GRB detected by HEND to date had a fluence of about 
$\rm 1.5 \times 10^{-6} erg\, cm^{-2}$.

\section{Verifying the IPN results}

Whenever a new spacecraft enters the network, both its timing and its ephemeris must
be verified in-flight.  There are various ways to do this, but the most effective end-to-end
test is to triangulate bursts whose positions are known with greater precision than that 
obtainable from the IPN alone.  This includes events (cosmic or SGR) localized with X-ray cameras,
and events with radio, X-ray, or optical counterparts.  To date, 12 GRBs which meet one
or more of these criteria, 5 bursts
from SGR1900+14, and 14 bursts from SGR1806-20 have been
used to verify \it Odyssey's \rm timing and ephemeris.  In these verifications, the 
triangulation is usually done using \it Odyssey \rm and a near-Earth spacecraft,
for which station-keeping is much simpler and more accurate, so that the timing and
ephemeris uncertainties for this spacecraft are negligible.

When such events are triangulated, a ``time shift'' can be derived, which is the amount of time which would
have to be added to the \it Odyssey \rm time to obtain perfect agreement between the annulus
derived from the \it Odyssey \rm data, and the known source position.  This time shift is therefore a measure
of the combination of statistical and systematic uncertainties in the triangulation.  The 
statistical uncertainty depends only on the statistics of the two time histories which
are being compared, and it is estimated in the cross-correlation analysis.  
The systematic uncertainty in these triangulations, which is a combination of timing and ephemeris errors,
is always found to be much less than the statistical one.  

\section{Some scientific highlights} 

\subsection{GRB020405}

This event was observed by \it Ulysses, Mars Odyssey, Konus, \rm and \it BeppoSAX \rm (the GRBM experiment).
The HEND and \it Ulysses \rm time histories are shown in figure 8.  The first localization, to a 75 square arcminute
error box, was produced within 16.6 hours (Hurley et al., 2002a).  Multi-wavelength observations
of the afterglow were carried out by numerous ground- and space-based telescopes.  An optical
transient was identified in the error box (figure 9), and a redshift of 0.695 was found
for it (Masetti et al. 2003).  A bump in the optical afterglow light curve advanced the 
case for a supernova/GRB connection (Dado, Dar, and de Rujula 2002; Masetti et al. 2003; Price et al. 2003).
Polarization studies of the optical afterglow were used to constrain the dust 
content of the host galaxy (Covino et al. 2002).  Masetti et al. (2003) achieved the 
first detection of an afterglow in the NIR bands, and the 
first detection of a galaxy responsible for an intervening absorption line system in the spectrum of
a GRB afterglow.  
Observations of a radio flare and of the the X-ray afterglow were 
used to deduce the nature of the surrounding medium (Berger et al. 2003;
Mirabal et al. 2003; Chevalier et al. 2004).  

\subsection{GRB020813}

This event was observed by \it Ulysses, Mars Odyssey, Konus, \rm, and HETE.  The HETE
WXM and SXC produced localizations (Villasenor et al. 2002) which were used by Fox, Blake, and
Price (2002) to identify the optical counterpart, whose redshift was measured to be 1.2 
(Price et al. 2002).  The IPN error box had an area
of 33 square arcminutes, and was circulated 24 hours after the burst (Hurley et al. 2002b,c).  It 
constrains the HETE positions, agrees with the 
optical position (figure 10), and serves both to verify them, and to confirm the
\it Odyssey \rm timing and ephemeris.  

\subsection{GRB021206}

This event was detected by \it Ulysses, Mars Odyssey, Konus, \rm RHESSI, 
and INTEGRAL (Hurley et al. 2002d, 2003).  Polarization
was detected in gamma-rays for this burst by RHESSI (Coburn and Boggs 2003; but see also
Rutledge and Fox, 2004 and Coburn and Boggs, 2004), and
the burst time history was used to set limits on the quantum gravity energy scale (Boggs et al. 2004).  This event
is a good example of the utility of the IPN technique.  The IPN as a whole has isotropic
response, and this burst occurred only 18 degrees from the Sun.  Thus it is unlikely that it could
have been detected and localized by INTEGRAL, HETE, or \it Swift \rm due to pointing constraints.  Since
RHESSI is the only spacecraft which has a demonstrated polarization measurement capability,
and it points in the solar direction, only IPN-localized bursts can benefit from
polarization measurements.  

\subsection{The giant flare from SGR1806-20}

An event which occurred on 2004 December 27 was the most intense burst
of X- and gamma-radiation detected to date (Hurley et al. 2005).  Although it was
observed by at least 20 spacecraft, none of them could localize it.  The precise
localization, which was a key element in identifying the source, was done by
triangulation using \it Odyssey\rm.  The burst turned out to be a giant flare from SGR1806-20.
This SGR was only about 5 \arcdeg from the Sun at the time of the burst.

\section{Conclusions}

The GSH and HEND experiments aboard the \it Mars Odyssey \rm spacecraft
have been successfully integrated into the third interplanetary network.  In this configuration, the
IPN is producing rapid, precise GRB localizations at the expected rate, and should
continue to do so for the forseeable future.  With the advent of newer missions, with much
greater sensitivities, and which are
capable of independently determining GRB positions to better accuracies and with
shorter delays, the question inevitably arises whether it is useful to maintain an IPN.
There are several arguments in favor of this.  The new missions achieve their
accuracy and sensitivity at the expense of sky converage; typically they view only 10 - 20\% of
the sky, while the IPN is isotropic.  Thus, first, IPNs serve as continuous monitors
of SGR activity throughout the galaxy, which the newer missions do not.  Second, IPNs will detect strong
bursts at a rate which is 5-10 times greater than that of the newer missions.  Therefore IPNs
should continue to serve a useful role in GRB and SGR studies for years to come.

\section{Acknowledgments}

We are grateful to John Laros for the initial work which made GRB detection possible
aboard Odyssey.  KH acknowledges support under the NASA Long Term Space Astrophysics 
and Odyssey Participating Scientist programs, FDNAG-5-11451.


\begin{figure}
\figurenum{1}
\plotone{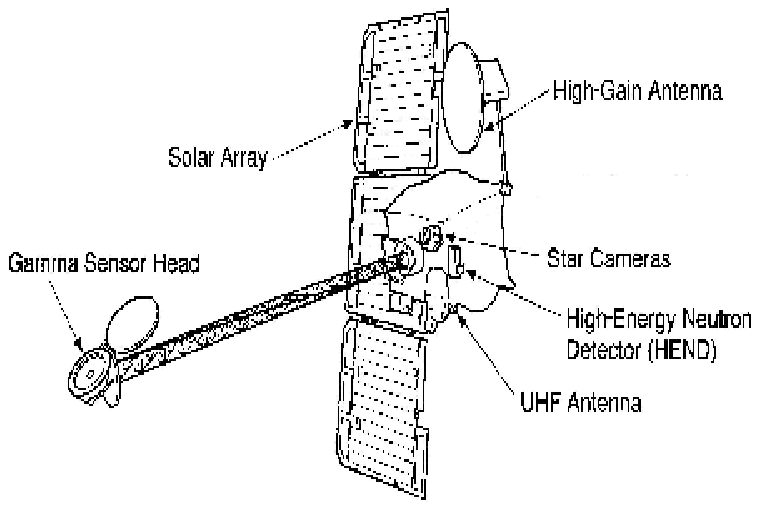}
\caption{The \it Mars Odyssey \rm spacecraft, showing the positions of the
HEND and GSH experiments, on the body of the spacecraft and on the boom,
respectively.  
}
\end{figure}


\begin{figure}
\figurenum{2}
\epsscale{0.8}
\plotone{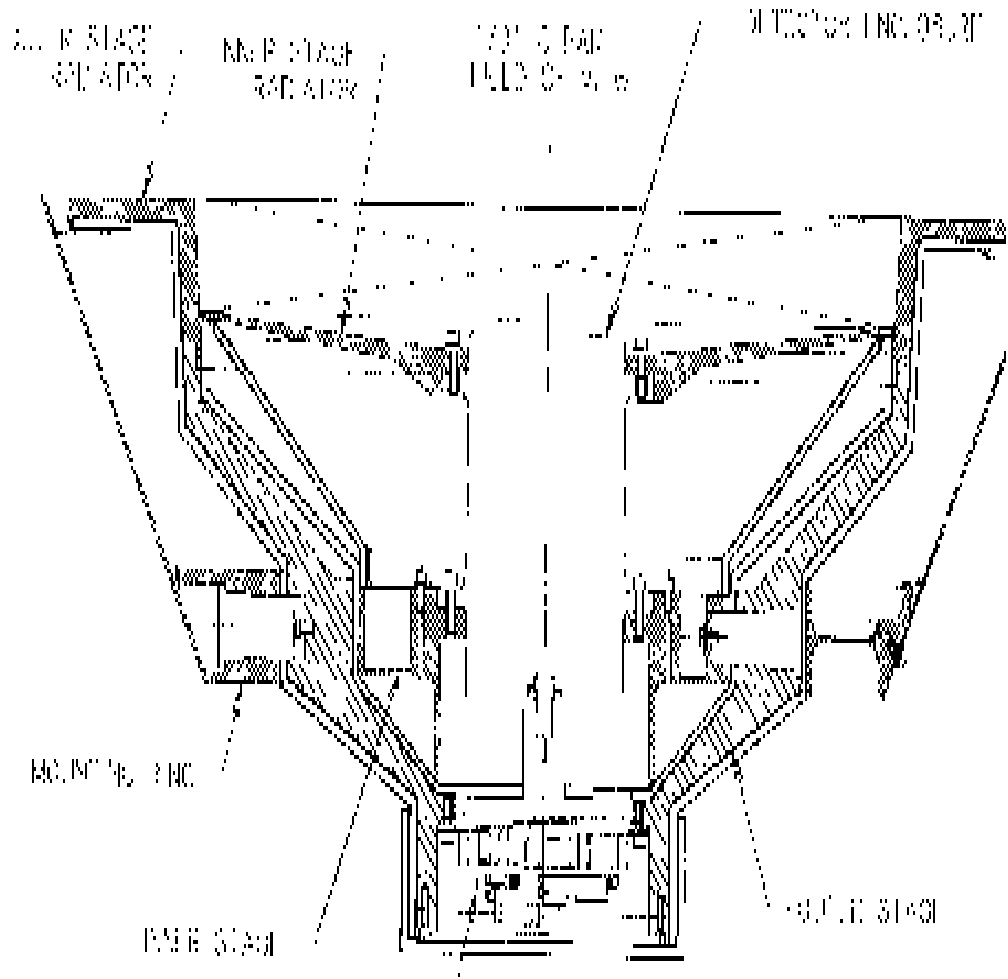}
\caption{Cross section of the gamma sensor head, showing the germanium
detector and the cooler.  The upper part of the head faces into space,
and Mars is toward the bottom.
}
\end{figure}


\begin{figure}
\figurenum{3}
\epsscale{0.5}
\plotone{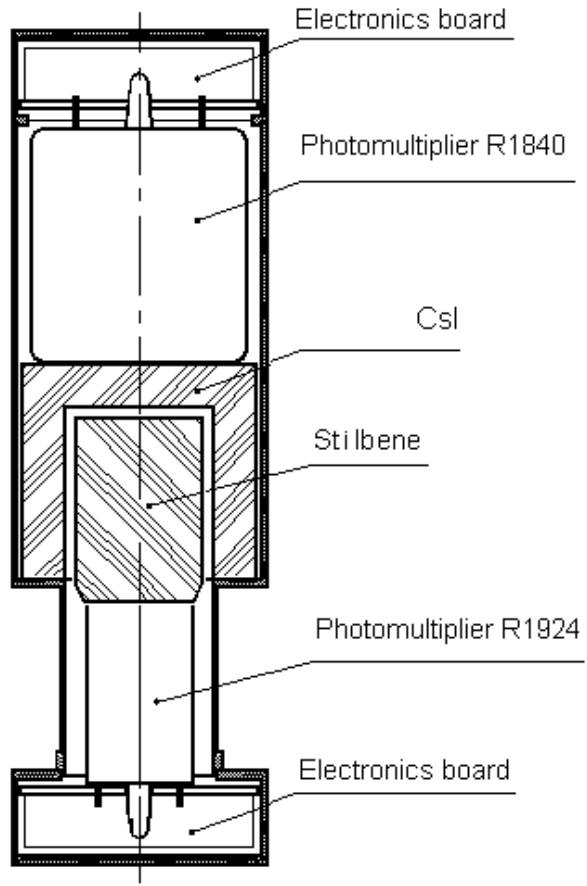}
\caption{The High Energy Neutron Detector (HEND) experiment.  The detector
axis points towards the top of the figure, and is indicated by the center line.
In the orbital phase of the mission, it points towards the nadir, that is,
towards the surface of Mars.  
}
\end{figure}


\begin{figure}
\figurenum{4}
\epsscale{0.6}
\plotone{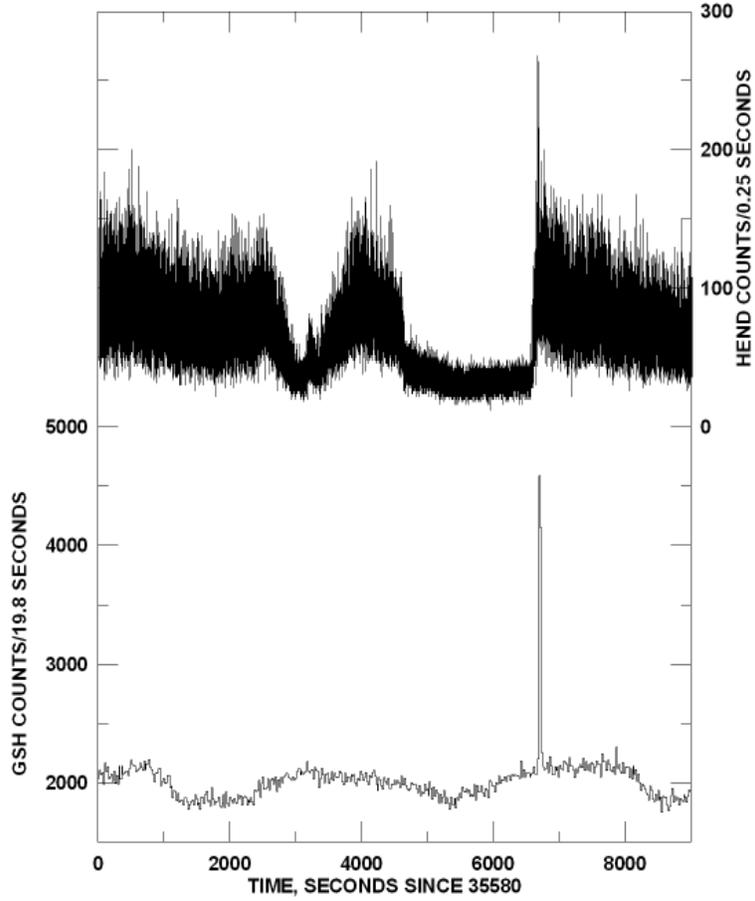}
\caption{A $\sim$ 9000 s (1.27 orbits) plot of the GSH (bottom, left hand axis) and HEND 
(top, right hand axis) counting rates on 2003 March 29, starting at 35580 s).
A bright gamma-ray burst occurred around 6600 s.}
\end{figure}


\begin{figure}
\figurenum{5}
\epsscale{0.8}
\plotone{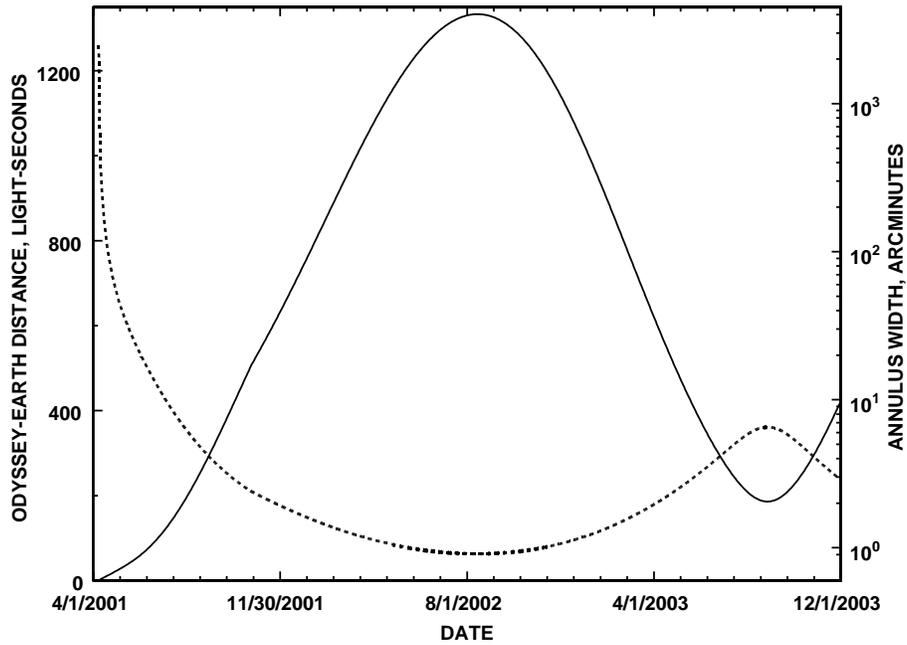}
\caption{Typical width of an IPN annulus obtained by triangulation between
\it Mars Odyssey \rm and a near-Earth spacecraft, as a function of time.
The left-hand scale shows the \it Odyssey \rm Earth distance in light-seconds
(solid line),
and the right-hand scale gives the annulus width in arcminutes (dotted line).  A cross-correlation
uncertainty of 200 milliseconds and a GRB arrival angle 30 $\degr$ from the Mars-Earth
vector have been assumed.  
}
\end{figure}


\begin{figure}
\figurenum{6}
\epsscale{0.6}
\plotone{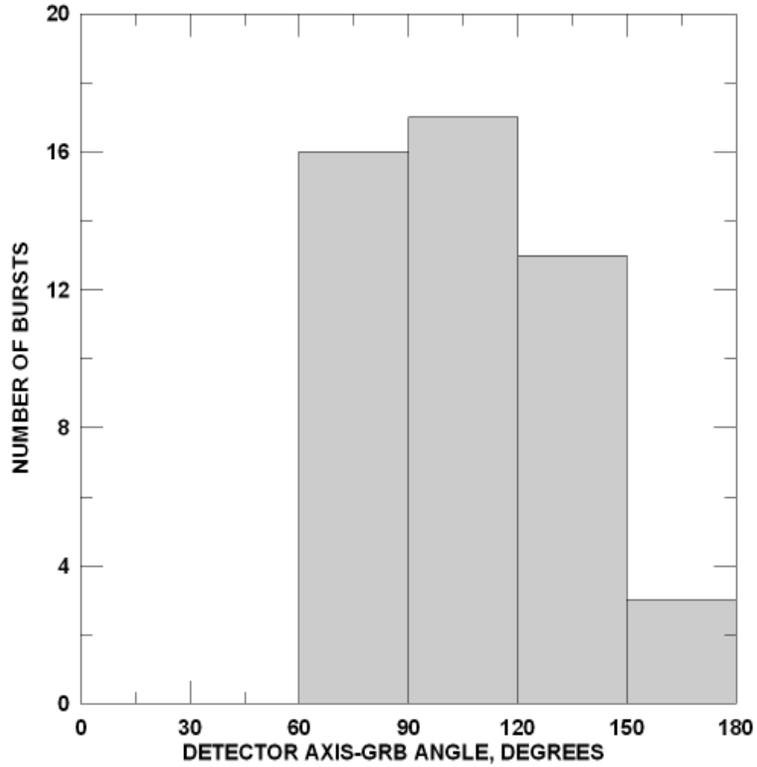}
\caption{The angular distribution of GRBs detected by HEND during the orbital
phase of the mission, relative to the detector axis.  
The detector axis is pointed towards Mars in the orbital phase, so no bursts are detected
at angles less than roughly 60 $\degr$.  Around 180 $\degr$, the relatively small detection 
cross-section (see figure 3) and spacecraft-blocking reduce the number of detections. 
}
\end{figure}

\begin{figure}
\figurenum{7}
\epsscale{0.6}
\plotone{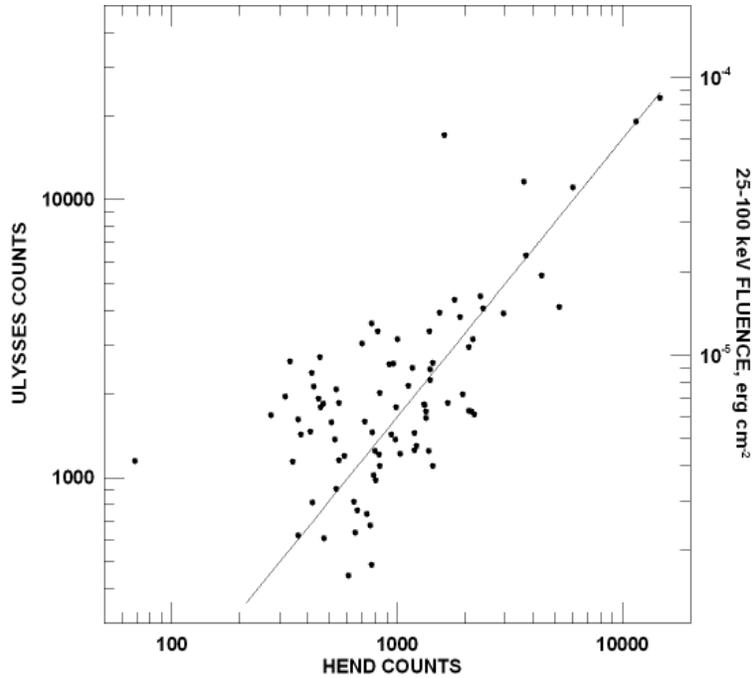}
\caption{The integrated counts for 87 GRBs detected by \it Ulysses \rm 
in the 25 - 150 keV energy range, versus the
integrated counts detected by HEND in the 30 - 1300 keV range.  The \it Ulysses \rm detector has an approximately
isotropic response, while the HEND cross-sectional area varies considerably 
depending on the arrival direction.  This, as well as the spectral differences of the
bursts, explains much of the
scatter in the points.  However, the HEND counts are approximately equal 
to the \it Ulysses \rm counts times a factor of 0.63 (straight line fit to the points).
The approximate fluences of the \it Ulysses \rm bursts are indicated.
}
\end{figure}


\begin{figure}
\figurenum{8}
\epsscale{1.}
\plotone{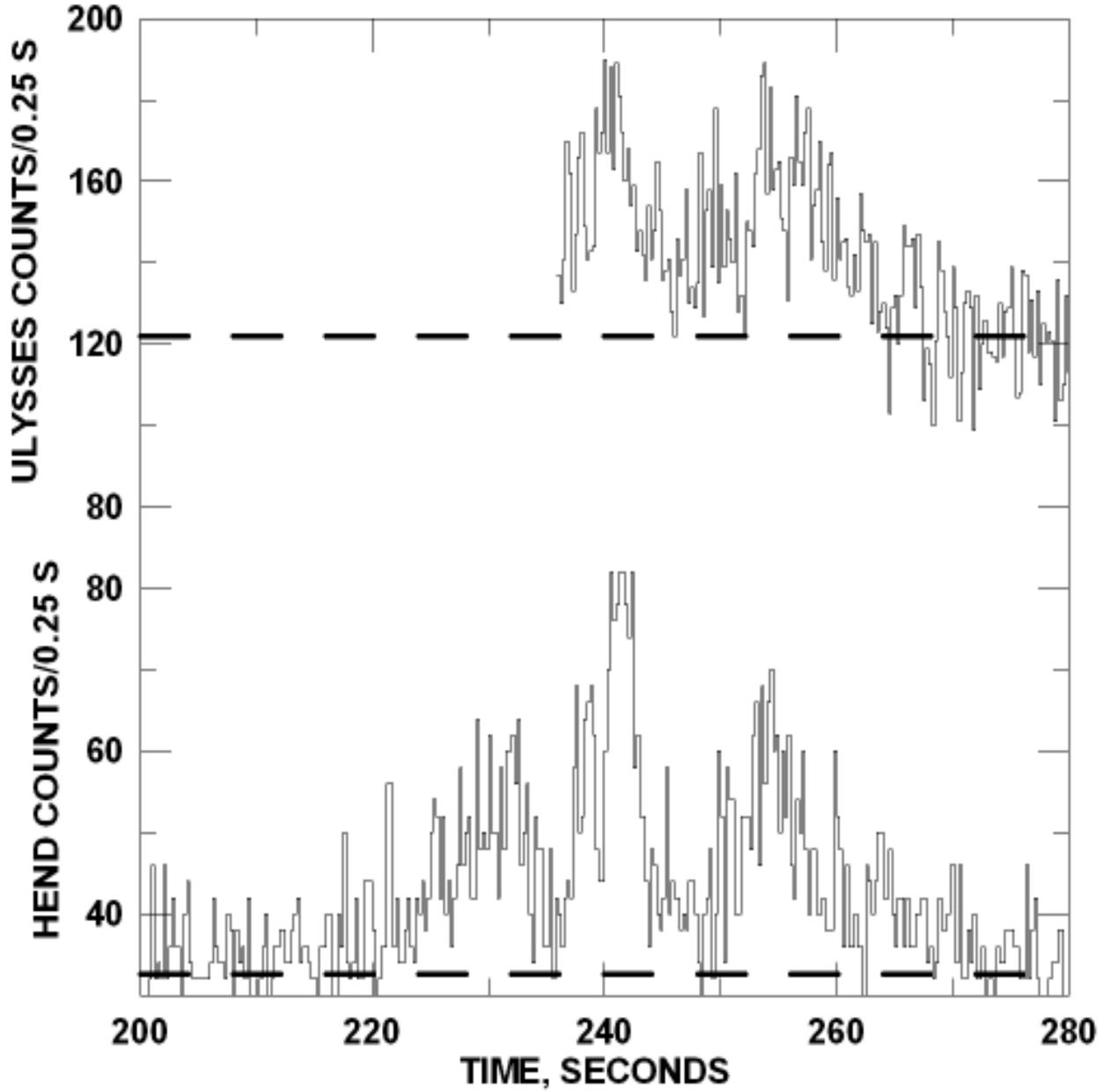}
\caption{The time histories of GRB020405, as detected by HEND and \it
Ulysses \rm.  \it Ulysses \rm triggered late in the burst.
}
\end{figure}


\begin{figure}
\figurenum{9}
\epsscale{0.8}
\plotone{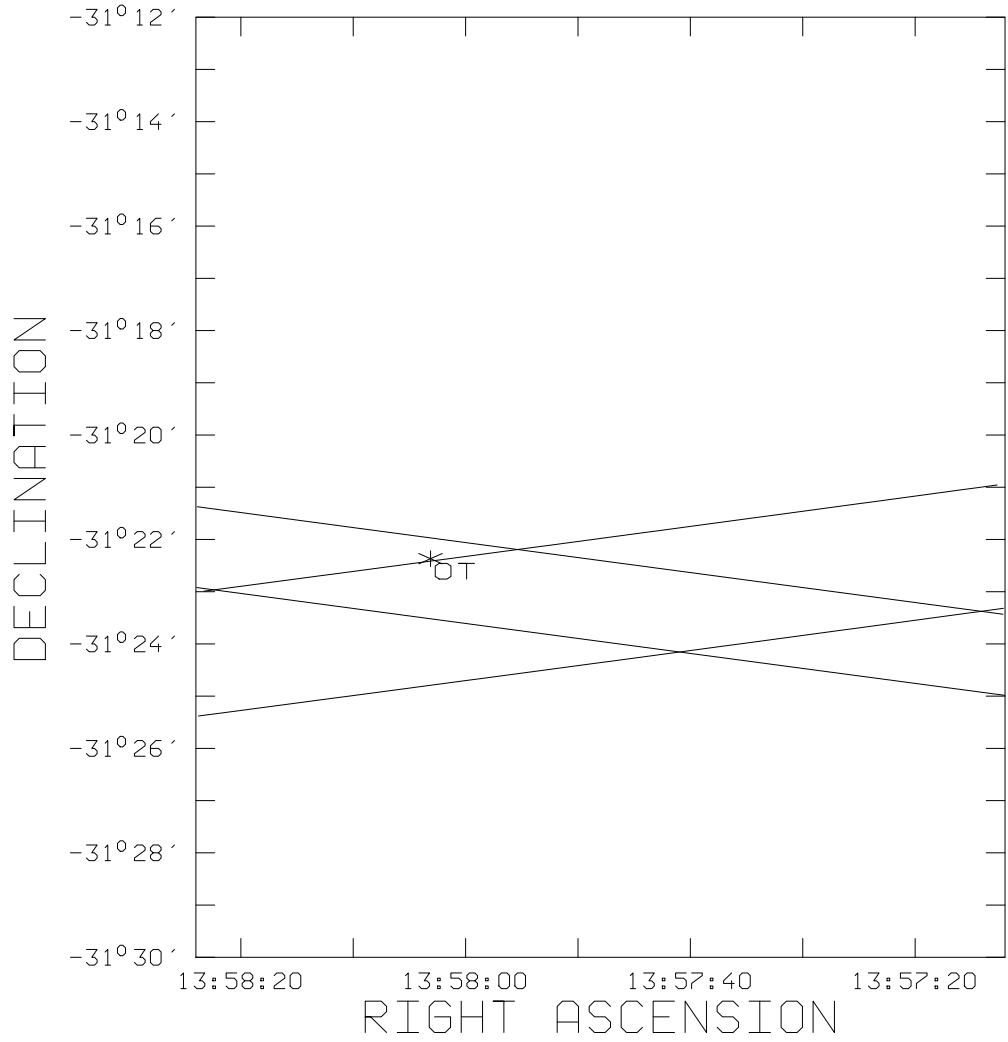}
\caption{The final, 14 sq. arcmin. IPN error box for GRB020405, defined by
the 2.3\arcmin wide \it Ulysses-Odyssey \rm and 1.5\arcmin wide \it Ulysses-BeppoSAX \rm annuli.
The position of the optical transient detected by
Price et al. (2003) is indicated.
}
\end{figure}


\begin{figure}
\figurenum{10}
\plotone{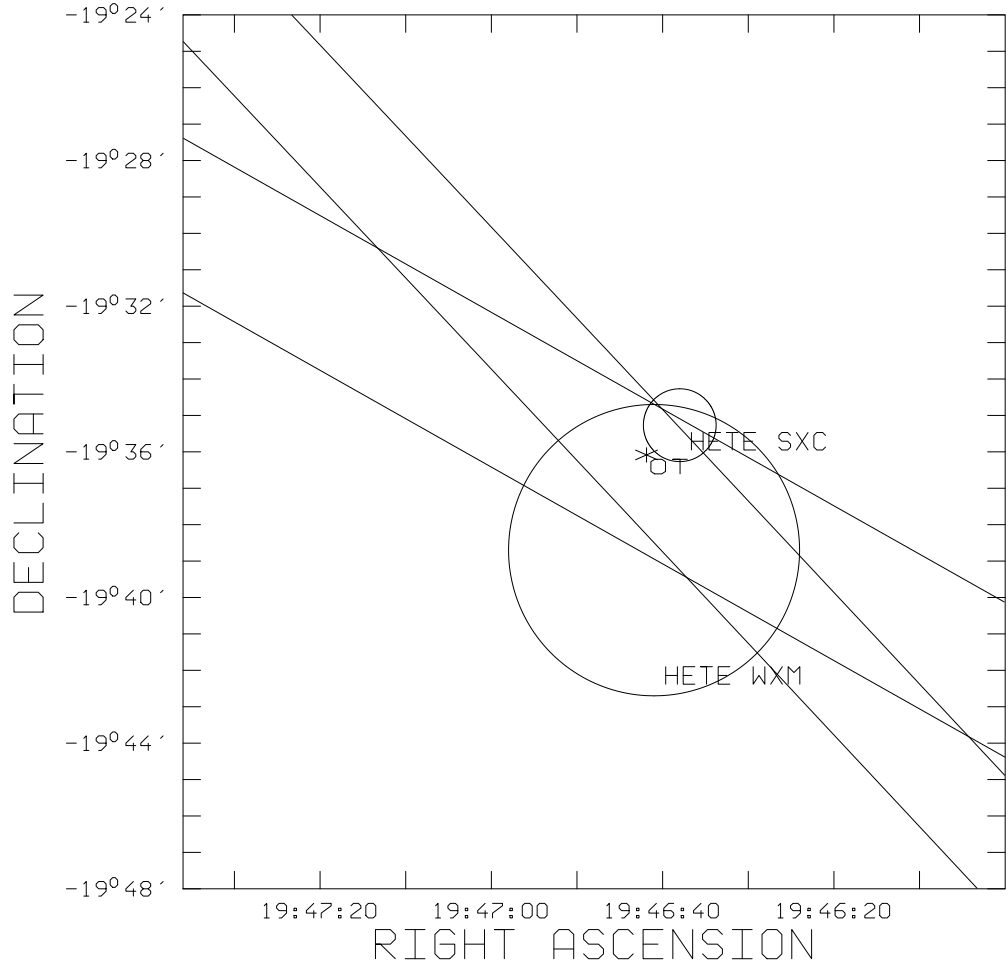}
\caption{IPN and HETE localizations of GRB020813. The IPN error box is defined by
the intersection of the 2.7 \arcmin wide \it Ulysses-Konus\rm \, annulus and the
3.7 \arcmin \it Ulysses-Odyssey\rm \, annulus.  The initial HETE SXC error circle
shown here was later revised by Jernigan et al. (2002) and agrees with the
position of the optical transient source (OT).
}
\end{figure}


\end{document}